# Study of transfer learning from 2D supercritical airfoils to 3D transonic swept wings

Runze Li,*  Yufei Zhang,†  Haixin Chen‡

(*Tsinghua University, Beijing, 100084, People's Republic of China*)

**Machine learning has been widely utilized in fluid mechanics studies and aerodynamic optimizations. However, most applications, especially flow field modeling and inverse design, involve two-dimensional flows and geometries. The dimensionality of three-dimensional problems is so high that it is too difficult and expensive to prepare sufficient samples. Therefore, transfer learning has become a promising approach to reuse well-trained two-dimensional models and greatly reduce the need for samples for three-dimensional problems. This paper proposes to reuse the baseline models trained on supercritical airfoils to predict finite-span swept supercritical wings, where the simple swept theory is embedded to improve the prediction accuracy. Two baseline models for transfer learning are investigated: one is commonly referred to as the forward problem of predicting the pressure coefficient distribution based on the geometry, and the other is the inverse problem that predicts the geometry based on the pressure coefficient distribution. Two transfer learning strategies are compared for both baseline models. The transferred models are then tested on the prediction of complete wings. The results show that transfer learning requires only approximately 500 wing samples to achieve good prediction accuracy on different wing planforms and different free stream conditions. Compared to the two baseline models, the transferred models reduce the prediction error by 60% and 80%, respectively.**

* Ph. D., School of Aerospace Engineering, email: lirz16@tsinghua.org.cn
†Associate professor, School of Aerospace Engineering, senior member AIAA, email: zhangyufei@tsinghua.edu.cn
‡ Professor, School of Aerospace Engineering, associate fellow AIAA, email: chenhaixin@tsinghua.edu.cn (Corresponding Author)

## Nomenclature

| | |
|---|---|
| AoA | = angle of attack |
| AR | = aspect ratio |
| **c** | = free stream condition vector |
| **C** | = wing planform parameter vector |
| $C_L$ | = lift coefficient |
| $C_p$ | = pressure coefficient |
| $\delta$ | = twist angle |
| **g** | = geometry parameter vector |
| $\lambda$ | = wing tapper ratio |
| $\Lambda$ | = swept angle |
| **p** | = pressure coefficient distribution vector |
| $M_\infty$ | = free-stream Mach number |
| Re | = Reynolds number |
| $t_{max}$ | = maximum relative thickness |
| $\theta$ | = dihedral angle |

## I. Introduction

Machine learning has attracted great attention in aerodynamic optimization and design because of its ability in high-dimensional nonlinear mapping, dimensionality reduction and pattern recognition [1]. Machine learning has been used for geometric filtering, aerodynamic coefficient prediction, flow field modeling, and machine learning-based optimizations [2]. Many machine learning algorithms have also been studied for various aerodynamic shapes, such as airfoils [3,4], blades [5,6], nacelles [7,8], and aircraft [9,10]. A thorough summary of the algorithms and applications can be found in [1] and [2].

Most machine learning studies are conducted on two-dimensional (2D) shapes and flow fields, as demonstrated in [1] and [2]. The main challenges for three-dimensional (3D) studies are the result of high dimensionality and high computational cost. Fig. 1 lists the approximate dimensions of conventional geometric parameterization methods for 2D airfoils, 3D wings and 3D aircraft. The sample sizes for machine learning are then approximated based on the geometric dimension and the examples in [2]. Usually, the data for machine learning are obtained by relatively expensive computational fluid dynamics (CFD) simulations. Therefore, the computational cost of 3D samples is much higher than the computational cost of 2D samples since both the number of simulations and the cost of one simulation are greater. Consequently, the machine learning of 3D wings and aircraft is much more difficult than the machine learning of 2D airfoils.

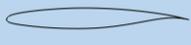

**Fig. 1 Features of machine learning for airfoils, wings and aircraft**

However, the flow field of airfoils can be regarded as the baseline of wings (sections). It is natural to consider using 2D data and models to help the aerodynamic coefficient prediction and the flow field modeling of 3D wings and aircraft. The gap between the 2D models and 3D flows is the 3D effect of the swept angle, the finite span, and the geometry changes in the spanwise direction, which brings nonnegligible changes to the baseline flow field and aerodynamic coefficients. The interference between wings, fuselage, pylons, and nacelles further brings a huge challenge for machine learning of real aircraft. Fortunately, there are simplified theories that describe some of the 3D effects, such as simple swept theory [11]. However, the other 3D effects still need to be learned.

Transfer learning is an approach to reuse a pretrained model on a new problem, which is currently very popular in deep learning because transfer learning can train models with comparatively little data [12,13]. An assumption of traditional machine learning is that the training data and testing data are taken from the same problem (i.e., the term 'domain' in transfer learning), so that the input and output spaces, as well as the data distributions, are the same [12]. When the training data are expensive or difficult to collect in a problem (domain), transfer learning is used to improve the model from this domain by transferring information from a related domain. For example, the traditional models are trained on airfoils and then used in the prediction of airfoils. Since there are often fewer samples available for 3D wings, transfer learning can be used to improve the performance of the airfoil model on the prediction of wing sections.

There are many machine learning applications in which transfer learning has been successfully applied, including text sentiment classification, image classification, and multilanguage text classification [12]. However, few studies have been carried out on aerodynamic optimization. Transfer learning was used mostly to transfer the model from low-fidelity data to high-fidelity data, e.g., from Data Compendium (DATCOM) results to CFD results [14], from XFOIL results to CFD results [15], or from coarse mesh CFD results to fine mesh results [16]. Transfer learning was also used in reinforcement learning [17,18], where the model to be transferred was used as the pretrained model for reinforcement learning. Transfer learning can be considered the pretraining process of reinforcement learning. To the best of our knowledge, there are no transfer learning studies to capture the 3D effects from 2D airfoils to 3D wings, especially in flow field modeling.

This paper proposes a physics-embedded 2D-to-3D transfer learning framework for swept supercritical wings, as shown in Fig. 2. The baseline models are trained on supercritical airfoil samples and reused in the prediction of swept wings through transfer learning. The transferred models are trained on a small number of wing samples. Simple swept theory is embedded in the framework to describe the general influence of the swept angle, and transfer learning is employed to capture the remaining 3D effects. Both the geometry prediction model and the pressure coefficient distribution prediction model are studied, and two transfer learning strategies are studied for both baseline models.

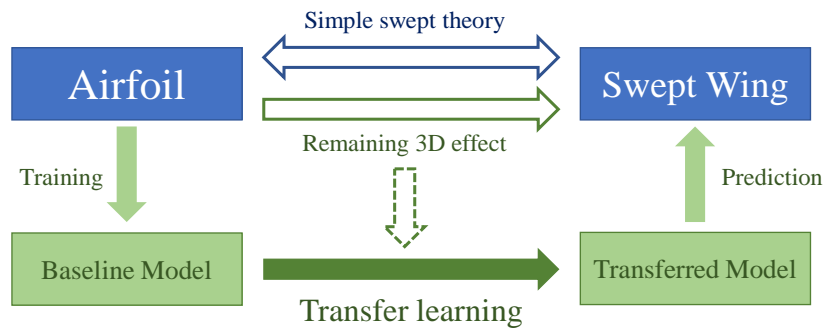

Fig. 2 Physics-embedded transfer learning from 2D airfoils to 3D swept wings

This paper is organized as follows. First, the airfoil samples and the swept wing samples are prepared. Next, the baseline models are trained on airfoil samples. The process of physics-embedded transfer learning and two transfer learning strategies are also introduced. Then, the performance of different transfer learning strategies and different sample sizes are studied. The transferred model is also utilized to predict the pressure coefficient contour and the geometry of swept wings. The results indicate that the transferred model can achieve good prediction accuracy on three-dimensional wings with an affordable number of samples.

## II. Sample preparation

### 2.1 Two-dimensional supercritical airfoils

The two-dimensional supercritical airfoils are built by the class shape transformation (CST) method [19] with a ninth-order Bernstein polynomial as the shape function. The airfoils have a unit chord length, and ten CST parameters are used to parameterize the upper surface and the lower surface. The Reynolds number (Re) based on the unit chord length equals $1 \times 10^7$. A C-grid is used for CFD with an open-source Reynolds averaged Navier-Stokes (RANS) solver, CFL3D [20]. The 301 grid points are distributed on the airfoil surface. The CFD settings and validation were introduced in our previous study [21], in which a representative airfoil sample set was generated. The range of free stream conditions and maximum thickness ($t_{max}$) of airfoil samples are listed in Table 1.

**Table 1 Ranges of free stream conditions and $t_{max}$**

|  | $M_\infty$ | $C_L$ | $t_{max}$ |
|---|---|---|---|
| Range | [0.71,0.76] | [0.60,0.90] | [0.09,0.13] |

The airfoil sample set contains 11,000 single shock wave supercritical airfoil samples, among which the free stream conditions and the $t_{max}$ are approximately uniformly distributed. The supercritical pressure distribution of airfoil samples is also designed to be as diverse as possible. Fig. 3 shows the geometry and pressure distribution of several typical airfoils under different conditions.

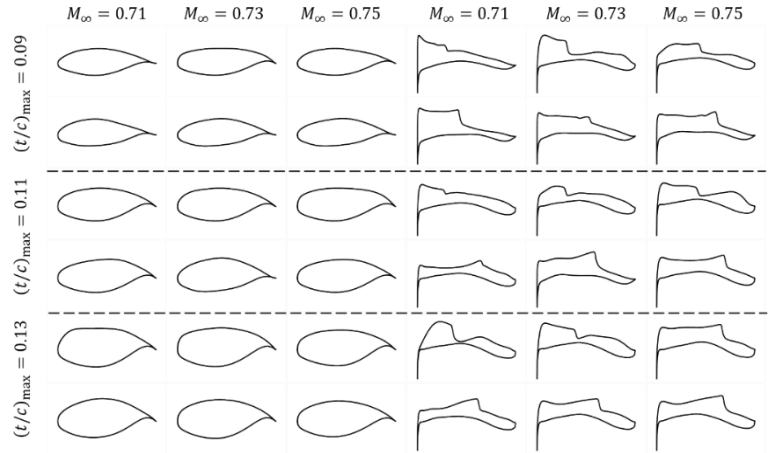

**Fig. 3 Geometries and pressure distributions of typical airfoil samples**

### 2.2 Three-dimensional simple wings

(1) Geometry parameterization method

The three-dimensional wings are studied using a wing-body configuration based on the NASA common research model (CRM) [22]. In this paper, simple wings without a kink are studied, where each wing surface is stretched from one baseline airfoil geometry. The free stream direction is defined as the positive X-direction, and the up direction is defined as the positive Y-direction. The fuselage is unchanged, and the chord length of the wing root section is kept at twelve meters. However, the baseline airfoil geometry and wing planform parameters are randomly sampled for different wings. The range of wing planform parameters is listed in Table 2. $\Lambda$ is the swept angle of the wing leading edge line, AR is the aspect ratio, $\lambda$ is the wing tapper ratio (i.e., the ratio of the wing tip chord length to the root chord length), and $\theta$ is the dihedral angle of the wing leading edge line. $\delta_{\text{root}}$ is the twist angle of the wing root section: the twist angle of the wing tip section ($\delta_{\text{tip}}$) always equals $-\delta_{\text{root}}$ in this paper, and the twist angle changes linearly from root to tip. The sign of $\delta_{\text{root}}$ is defined by the right-hand rule on the Z-axis. $r_{\text{t,root}}$ is the ratio of the relative maximum thickness of the wing root section to the baseline airfoil $t_{\max}$. $r_{\text{t,t/r}}$ is the ratio of the relative maximum thickness of the wing tip section to the wing root section. Fig. 4 shows the reference aircraft in gray shades, and three typical wing planforms are demonstrated with black lines.

**Table 2 Ranges of wing planform parameters**

|  | $\Lambda$ (°) | AR | $\lambda$ | $\theta$ | $\delta_{\text{root}}$ | $r_{\text{t,root}}$ | $r_{\text{t,t/r}}$ |
|---|---|---|---|---|---|---|---|
| Range | [0,40] | [6,14] | [0.3,1.0] | [0,30] | [−3,0] | [0.9,1.1] | [0.7,1.0] |

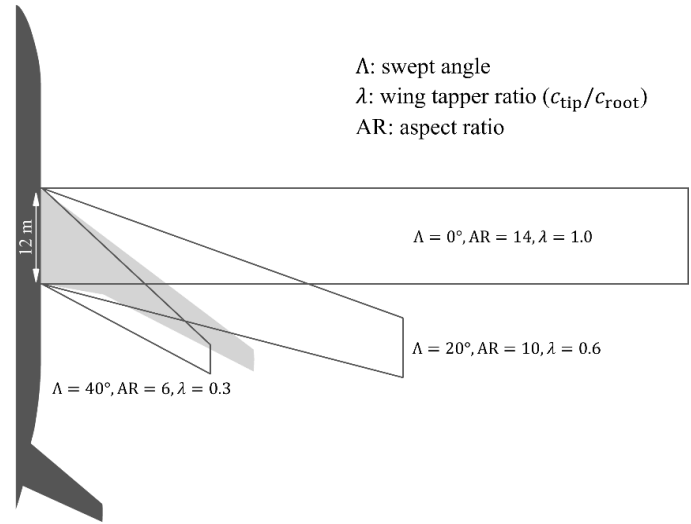

**Fig. 4 The reference aircraft and three typical wing planforms**

(2) CFD method

The BLWF code is a fast aerodynamic analysis tool for transonic aircrafts developed at the Central Aerohydrodynamic Institute (TsAGI), which is based on an iterative quasisimultaneous algorithm of strong viscous-inviscid interaction of external flow and boundary layer on lifting surfaces. The flow field can be solved within a minute on a modern personal computer (PC) [23,24]. A lower version of BLWF is available for noncommercial use on the official website (http://blwf-aero.ru/index_en.html), which is used in this paper. The BLWF is validated with CRM model experiments [25] and CFL3D results, as shown in Fig. 5. The pressure coefficient ($C_p$) contour of the CFL3D result is also plotted in Fig. 5. The locations of the five sections are plotted on the contour.

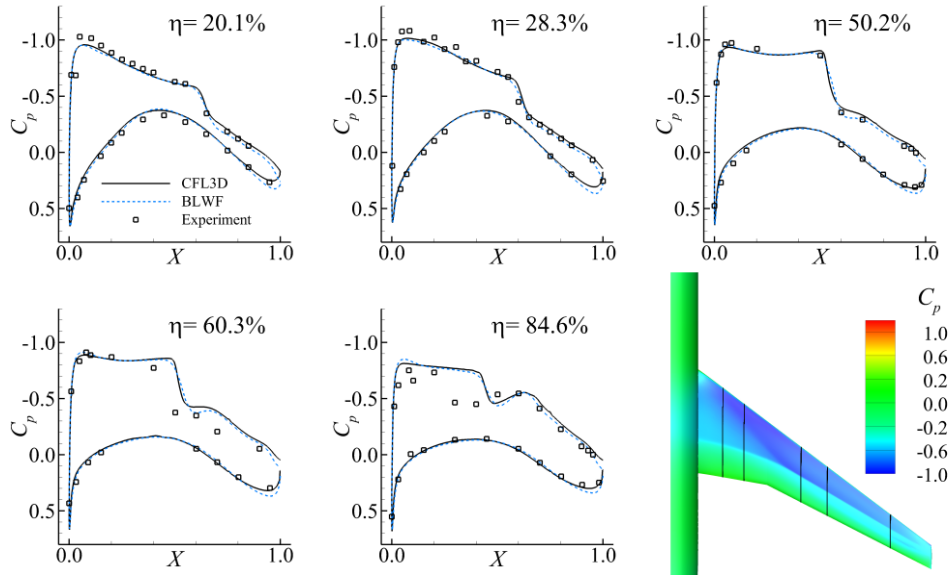

**Fig. 5 Wing section pressure distributions of CFL3D, BLWF, and experimental results**

The main reason for utilizing BLWF for the CFD simulations is that numerous extra wing samples are required for the model validation in this study. The following study will show that a few hundred wing samples are sufficient for transfer learning to make two-dimensional models achieve satisfying performance in three-dimensional predictions. It is quite affordable to use Reynolds-averaged Navier-Stokes (RANS) solvers for their CFD simulations. However, more than 10,000 wing samples are generated to fully validate the transfer learning strategies in this study, so using BLWF is the only possible choice.

Although BLWF is not as accurate as RANS solvers such as CFL3D in transonic flow simulations, especially when the shock wave boundary layer interaction is severe, BLWF produces reasonable results for most of the samples in this study. Furthermore, the relatively lower accuracy of BLWF simulations does not affect the correctness and generality of the transfer learning studies in this paper. The effects of finite span, swept angle, spanwise geometry variation, etc., are the major contributors to the difference between two-dimensional airfoils and three-dimensional wings. Therefore, RANS solvers can be used for future practical applications of transfer learning, and the effectiveness of transfer learning will not be damaged.

(3) Sampling of wings

An iterative sampling process is conducted because the degree of freedom (DoF) of wings is too high, and it is difficult to generate reasonable wings with simple random sampling. As shown in Fig. 4, simple wings without a kink are studied, where each wing surface is stretched from one baseline airfoil geometry. For each wing sample, its baseline airfoil is randomly selected from the airfoil sample set described in Section 2.1, and its wing planform parameters, i.e., $\{\Lambda, AR, \lambda, \theta, \delta_{\text{root}}, r_{\text{t,root}}, r_{\text{t,t/r}}\}$, are randomly sampled. The thickness ratios $r_{\text{t,root}}$ and $r_{\text{t,t/r}}$ make the relative thickness of wing sections different from the baseline airfoil, thereby introducing additional variation to the wing geometry. The free stream condition is also randomly sampled from the ranges listed in Table 3. The ranges of free stream Mach number $M_{\infty,\text{wing}}$ and angle of attack $\text{AoA}_{\text{wing}}$ cover most flight conditions encountered by civil transonic aircraft. The Reynolds number is calculated with its $M_{\infty,\text{wing}}$, the chord length of the wing root section, and the air properties at an altitude of 10 km.

**Table 3 Ranges of wing free stream conditions**

|  | $M_{\infty,\text{wing}}$ | $\text{AoA}_{\text{wing}}(°)$ |
| --- | --- | --- |
| Range | [0.68,0.78] | [0,4] |

## 2.3 Wing section samples for transfer learning

Transfer learning is applied to models trained on two-dimensional airfoils, and the input and output of these models are airfoils. Therefore, the transferred model directly addresses wing sections instead of three-dimensional wing surfaces. The dashed lines in Fig. 6 are some examples of wing sections. After the model is trained, the pressure coefficient contour on the wing surface, or the wing geometry, can be obtained by predicting and interpolating multiple wing sections so that the transferred two-dimensional model can be used for the prediction of three-dimensional wings. In this paper, 21 sections evenly distributed in the range of 20%-90% semispan are used to construct the pressure coefficient contour.

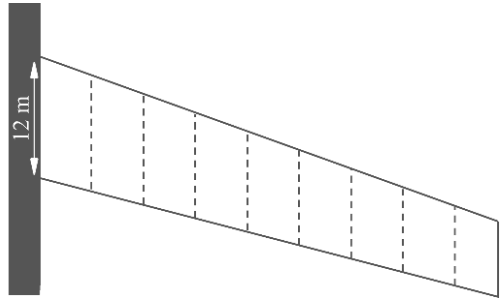

**Fig. 6 Wing section examples (dashed lines)**

Wing section samples are needed for transfer learning. Multiple sections are extracted from each wing sample. The spanwise location of sampled sections is not limited to the examples in Fig. 6. Then, the wing sections are transformed for transfer learning based on the simple swept theory, as shown in Fig. 7. The wing section thickness, geometric parameters, pressure coefficients, free stream Mach number and lift coefficient are transformed to their equivalent 2D values. The transformed wing section samples are used for transfer learning to better capture the other 3D effects shown in Fig. 2. Then, the simple swept theory is applied again to convert the equivalent 2D values back to the actual 3D values when the transferred model is used to predict 3D wings.

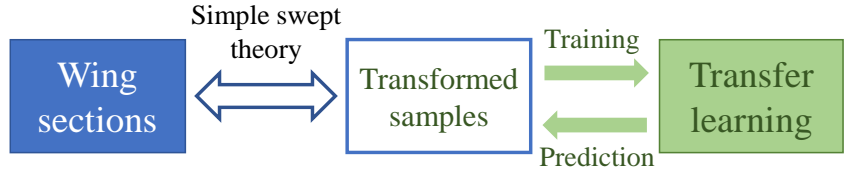

**Fig. 7 The utilization of the simple swept theory in transfer learning**

The simple swept theory [11] is:

$$M_{\infty,\text{section}}^{\text{3D}\rightarrow\text{2D}} = M_{\infty,\text{wing}} \times \cos\Lambda$$
$$t_{\text{section}}^{\text{3D}\rightarrow\text{2D}} = t_{\text{section}}/\cos\Lambda \quad (1)$$
$$C_{p,\text{section}}^{\text{3D}\rightarrow\text{2D}} = C_{p,\text{section}}/\cos^2\Lambda$$

, where $M_\infty$ is the free stream Mach number, $t$ is the thickness, $C_p$ is the pressure coefficient, and $\Lambda$ is the swept angle. The superscript '3D→2D' denotes the equivalent 2D value transformed from 3D values.

Several constraints are applied to the wing sections during sampling so that reasonable samples are generated.

1) The equivalent free stream Mach number of wing sections ($M_{\infty,\text{section}}^{\text{3D}\rightarrow\text{2D}}$) should be within the range of [0.68,0.78], which is slightly wider than the range of airfoil samples ([0.71,0.76]).

2) The equivalent relative maximum thickness of wing sections ($t_{\max,\text{section}}^{\text{3D}\rightarrow\text{2D}}$) should be within the range of [0.08,0.14], which is slightly wider than the range of airfoil samples ([0.09,0.13]).

3) The equivalent lift coefficient of wing sections ($C_{L,\text{section}}^{\text{3D}\rightarrow\text{2D}}$) should be within the range of [0.3,0.9], which is wider than the range of airfoil samples ([0.6,0.9]). $C_{L,\text{section}}^{\text{3D}\rightarrow\text{2D}}$ is the integral of $C_{p,\text{section}}^{\text{3D}\rightarrow\text{2D}}$ (Eq. (1)) on the X-axis ($X \in [0,1]$). Although $C_{L,\text{section}}^{\text{3D}\rightarrow\text{2D}}$ does not exactly match the definition of the lift coefficient, the difference can be learned by transfer learning and will not damage the performance of the transferred model.

The sampling process is described in Algorithm A, in which 5,000 valid wings are generated based on the airfoil sample set in Section 2.1. Three to five wing section samples are extracted from each wing, and 10,000 wing sections are randomly sampled from the first 4,500 valid wings. The remaining 500 wings are used for the future testing of the 3D wing prediction.

Algorithm A: Sampling of wings and wing sections

| | |
|---|---|
| 1 | Number of valid wing samples $N_{\text{wing}} = 0$; |
| 2 | while ($N_{\text{wing}} \leq 5{,}000$); { |
| 3 |   Randomly sample an airfoil from the airfoil sample set in Section 2.1; |
| 4 |   Randomly sample the wing planform parameters, i.e., $\{\Lambda, \text{AR}, \lambda, \theta, \delta_{\text{root}}, r_{\text{t,root}}, r_{\text{t,t/r}}\}$; |
| 5 |   if (the sampled wing exists in the sample set) {continue}; |
| 6 |   Run CFD simulation of the sampled wing with BLWF; |
| 7 |   if (BLWF converges); { |
|  |     Randomly select five spanwise locations $Z_{\text{section}} \in [0.2\, Z_{\text{tip}}, 0.9\, Z_{\text{tip}}]$; |
| 8 |     ($Z_{\text{tip}}$ is the Z coordinate of the wing tip.) |
| 9 |     Extract the geometry and pressure distribution of these sections; |
| 10 |     if (more than three sections satisfy the aforementioned constraints); { |
| 11 |       $N_{\text{wing}} = N_{\text{wing}} + 1$; |
| 12 |       Add the valid sections to the wing section sample set; |
| 13 |     } end if; |
| 14 |   } end if; |
| 15 | } end; |

## III. The baseline models of two-dimensional airfoils

The baseline models predicting the airfoil pressure coefficient distribution and the airfoil geometry are introduced and trained in this section.

### 3.1 Architecture of the baseline model

Two types of predictive models are often utilized in aerodynamic optimizations and fluid dynamic studies.

(1) forward problem: predicting the surface pressure coefficient distribution ($\boldsymbol{p}$) or the entire flow field based on the geometry ($\boldsymbol{g}$) and the free stream condition ($\boldsymbol{c}$), i.e., $(\boldsymbol{g}_0, \boldsymbol{c}_0) \to \boldsymbol{p}$, as shown in Fig. 8 a);

(2) inverse problem: predicting geometry based on the provided pressure coefficient distribution and the free stream condition, i.e., $(\boldsymbol{p}_0, \boldsymbol{c}_0) \to \boldsymbol{g}$, as shown in Fig. 8 b).

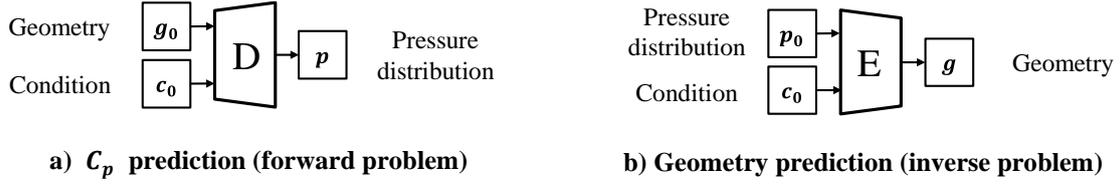

a) $C_p$ prediction (forward problem)  b) Geometry prediction (inverse problem)

**Fig. 8 Architecture of the baseline models**

The subscript 0 means the value is the ground truth value. The dimension of the airfoil pressure coefficient distribution ($p$) is 401. $g$ is the CST parameter of the airfoil upper and lower surfaces, of which the dimension is 20.

A conditional encoder-decoder architecture is used to train both types of predictive models at the same time, as shown in Fig. 9. The decoder is the first type of predictive model, and the encoder is the second. The dashed lines show the mean square error (MSE) losses for training. The final loss function of the entire model is:

$$\text{loss} = \text{MSE}(p, p_0) + r \cdot \text{MSE}(g, g_0) \qquad (2)$$

, where $r$ is the weight of the encoder loss, $r = 0.1$. The pressure coefficient $p$ and CST parameters $g$ are normalized to $[-1, 1]$ by their maximum and minimum values in the airfoil sample set.

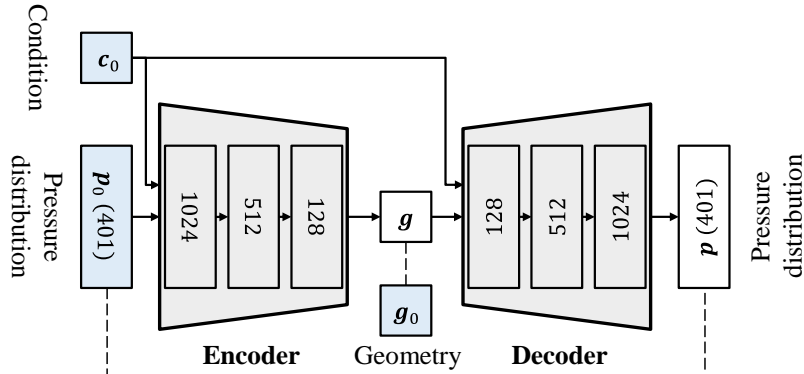

**Fig. 9 Architecture of the conditional encoder-decoder**

The model is trained and tested on the airfoil sample set described in Section 2.1, where 10,000 airfoils are randomly selected as the training set, and the remaining 1,000 airfoils are used for the test set. The Adam optimizer is used for training. The initial learning rate is 0.0002, which is reduced by a factor of 0.1 every 1,000 steps. The model is trained by 10,000 epochs, where the minibatch size is 64. Fig. 10 shows the history of loss functions on the training set and the test set. The blue lines show the decoder loss, i.e., the first term on the right-hand side of Eq. (2). The green lines show the encoder loss. The lines without scatters are the results on the training set, showing that both the encoder and the decoder have prediction accuracy similar to the training set and the test set.

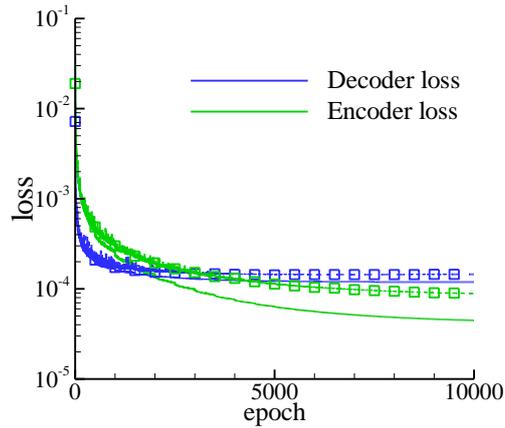

**Fig. 10 Training history of the two-dimensional model**

### 3.2 Testing the baseline model on airfoils and wing sections

Fig. 11 shows some examples of airfoils in the test set. The first row shows the performance of the decoder, and the second row shows the performance of the encoder. The encoder predicts the CST parameters ($g$) based on the pressure coefficient distribution input ($p_0$). The airfoil geometries shown in Fig. 11 are generated based on the predicted CST parameters. The black solid lines in Fig. 11 are the ground truth data, and the dashed lines are the predicted data. The first column shows the airfoils with the median value of the root mean square error (RMSE) in the test set, the second column shows the airfoils with the 90$^{th}$ percentile of RMSE, and the third column shows the airfoils with the maximum RMSE. The '90$^{th}$ percentile of RMSE' means that 90% of samples have a smaller RMSE than the 90$^{th}$ percentile value. The mean RMSE of the decoder is 3.22%, and the mean RMSE of the encoder is 0.36%. The results show that both the decoder and encoder achieve good performance on most of the examples. The models can be used as the baseline models for further transfer learning.

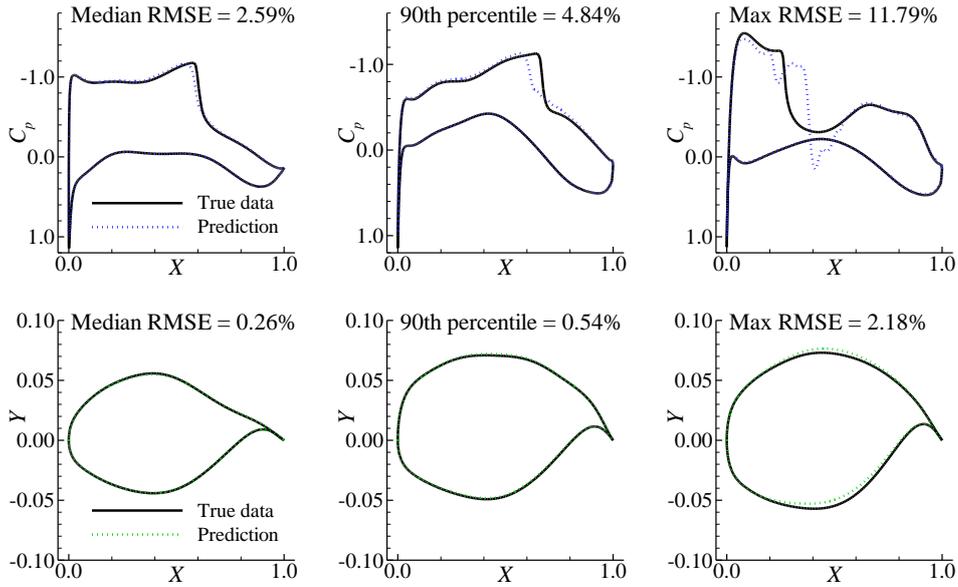

**Fig. 11 Examples of airfoils with different prediction accuracies in the test set**

The baseline models are tested on the transformed wing section samples from Section 2.3. The transformed samples are obtained by applying the simple swept theory on the wing sections, as shown in Fig. 7. Therefore, the 3D effect of the finite span, the wing-body interference, the geometry change in the span direction, etc., have yet to be learned. Fig. 12 shows the performance of the baseline model on the transformed samples, which, as expected, is much worse than its performance on the airfoil sample set. The mean RMSE of the decoder is 10.30%, and the mean RMSE of the encoder is 4.94%. Therefore, it is necessary to improve their performance on 3D wing sections.

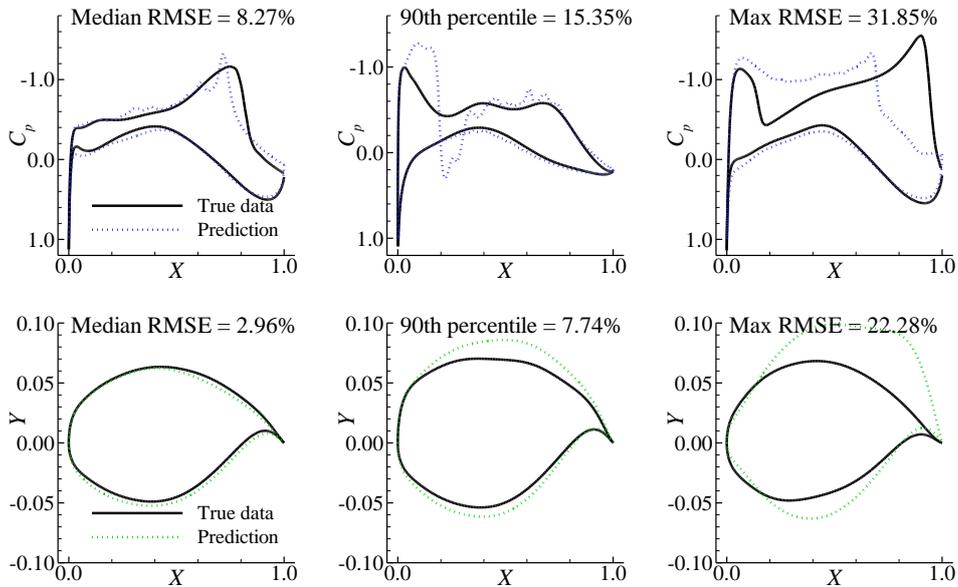

**Fig. 12 Examples of transformed wing sections with different prediction accuracies**

# IV. Transfer learning strategies and results

The process of transfer learning from 2D airfoils to 3D swept wings was summarized. Then, two transfer learning strategies were studied for the airfoil geometry prediction model and the pressure coefficient distribution prediction model. The transfer learning process is demonstrated in Fig. 13:

(1) The baseline models are trained on airfoil samples, as shown in Section 3.1;

(2) The swept wing samples are constructed from the airfoil samples, and the wing sections are then transformed based on the simple swept theory, as introduced in Sections 2.2 and 2.3.

(3) The baseline models are reused and improved on the transformed wing sections by different transfer learning strategies;

(4) The transferred models are used to predict transformed wing sections that are evenly distributed in the wingspan. Then, the predicted sections are transformed back to the actual wing sections so that the complete wing can be constructed.

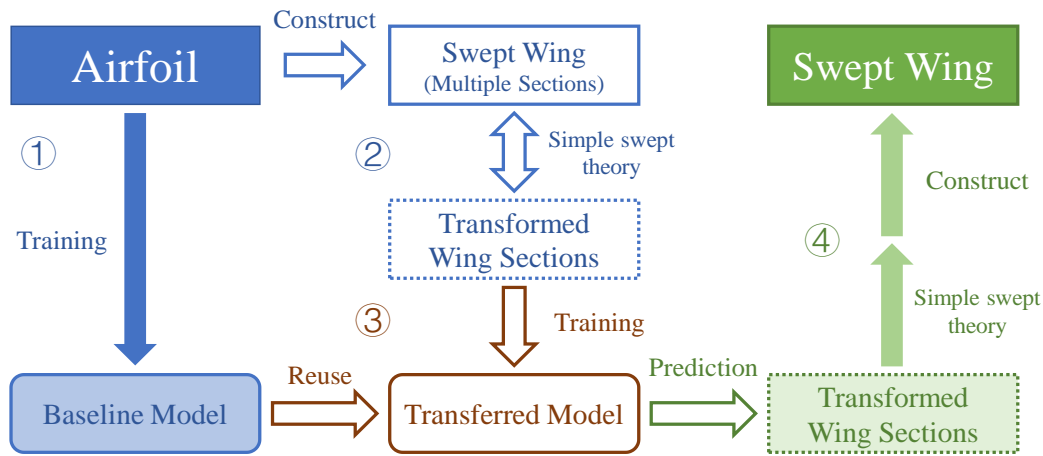

**Fig. 13 Process of 2D-to-3D transfer learning**

## 4.1 Transfer learning process and strategies

This section introduces the strategies of transfer learning studied in this paper, which is included in the third step in Fig. 13. Transfer learning is usually divided into homogeneous and heterogeneous transfer learning [12]. Homogeneous transfer learning handles the situations where the domains are of the same feature space, which means the input and output spaces are the same for the baseline model and the transferred model. The marginal distribution difference or the conditional distribution difference between domains needs to be learned during transfer learning [12]. Transfer learning strategies can also be categorized into four groups in other manners:

instance-based, feature-based, parameter-based, and relation-based approaches. The detailed definition and examples can be found in Ref. [13].

Parameter-based transfer learning, which transfers knowledge at the model/parameter level [13], is utilized in this paper. The parameters of the baseline model are fixed. An additional neural network is appended to the baseline model, which is trained to improve the prediction accuracy on the new problem. In this way, the knowledge contained in the pre-obtained baseline models can be transferred into the transferred model during the training process.

The general process of 2D-to-3D transfer learning is demonstrated in Fig. 14. The baseline models trained in Section 2.1 map the relationship between the airfoil geometry ($g$) and pressure coefficient distribution ($p$). Simple swept theory is embedded to capture the general influence of the swept angle. Then, the additional neural network (denoted by NN) is trained to capture the remaining 3D effects caused by the finite span, spanwise geometry change (e.g., tapper ratio, thickness change), and wing-body interference.

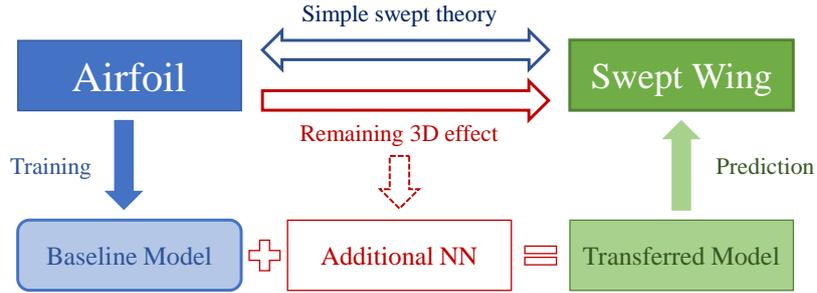

**Fig. 14 Process of physics embedded parameter-based transfer learning**

Two baseline models (Fig. 8) are studied, i.e., the pressure coefficient distribution prediction mode (the baseline decoder, denoted by D) and the airfoil geometry prediction model (the baseline encoder, denoted by E). Two parameter-based transfer learning strategies are established for both baseline models. The model architectures are shown in Fig. 15 and Fig. 16. The additional neural network (NN) is a fully connected network with two or three middle layers in all four strategies, where each middle layer has 1,024 neurons.

$g$ in Fig. 15 and Fig. 16 are the geometric parameters of an airfoil or a wing section, $p$ is the pressure coefficient distribution, $c$ contains the free stream conditions ($M_\infty, C_L$) of an airfoil or the transformed value of a wing section, and $C$ contains the wing planform parameters, i.e., $\{\Lambda, \text{AR}, \lambda, \theta, \delta_{\text{root}}, r_{\text{t,root}}, r_{\text{t,t/r}}\}$. The tilde (~) over characters marks the variables that are transformed by the simple swept theory. The derivative notation ($'$) marks the intrinsic variables that are fed into other models.

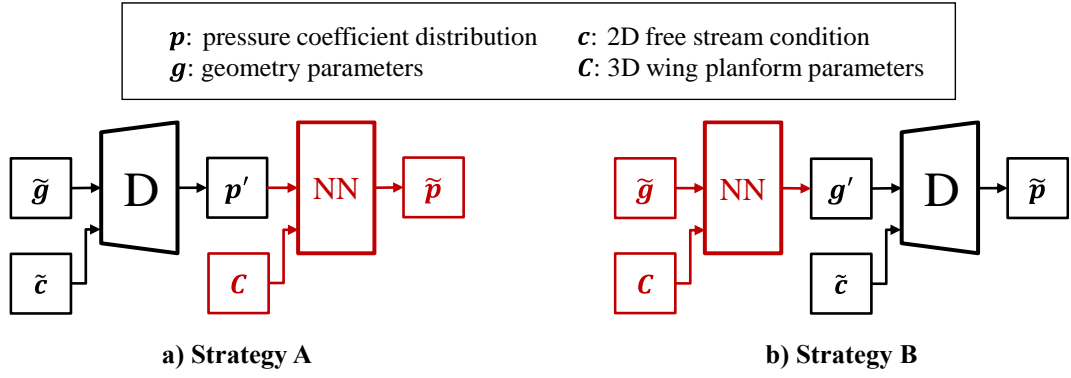

a) Strategy A  b) Strategy B

**Fig. 15 Model architectures of the two transfer learning strategies for the forward problem (Model D)**

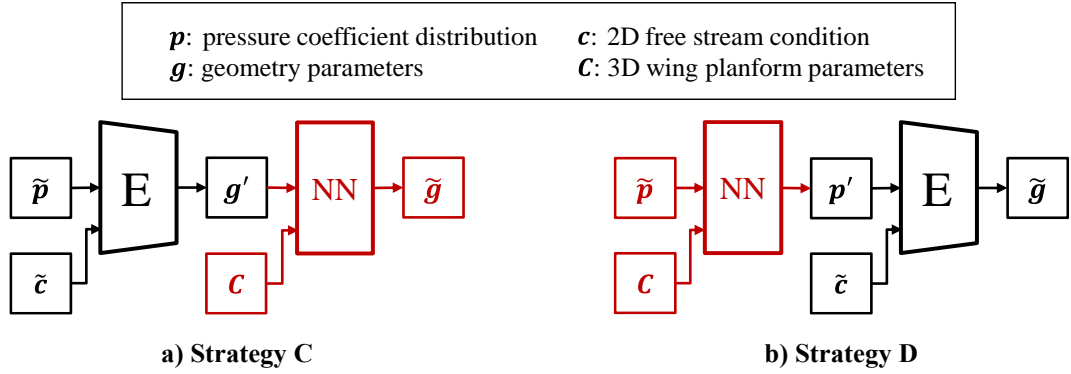

a) Strategy C  b) Strategy D

**Fig. 16 Model architectures of the two transfer learning strategies for the inverse problem (Model E)**

### 4.2 Influence of the training set size on transfer learning

The four transfer learning strategies are trained and tested on the 10,000 transformed wing section samples in Section 2.3. Each strategy is trained four times on different sizes of the training set, which are 100, 200, 500, 1,000, 2,000, 5,000, and 8,000. The rest of the samples are used as the test set. The Adam optimizer is used for training. The initial learning rate is 0.0001, which is reduced by a factor of 0.1 every 400 steps. The model is trained by 2,400 epochs. The minibatch size is 8, 16, 16, 32, 64, 128, and 256 for the seven different training set sizes.

The prediction errors of the four strategies on different training set sizes are plotted in Fig. 17 and Fig. 18. The solid lines are the RMSE on the training samples, and the dashed lines are the RMSE on the test samples. The black lines are the average RMSE, the red lines are the median value of RMSE, and the green lines are the $90^{th}$ percentile value of RMSE. The $90^{th}$ percentile value means that 90% of the samples have a smaller RMSE than the $90^{th}$ percentile value. The scatters show the mean value of four runs for each strategy, and the error bar shows the standard deviation of four runs.

Since the dimensions of $p$ and $g$ are different, the influence of the size of neural networks needs to be discussed. Additional networks with two or three middle layers are trained separately for each strategy, and each strategy is trained twice. The standard deviation of RMSE in the four runs of each strategy is shown by the error bars in Fig. 17 and Fig. 18. The training set and the test set are randomly divided in each run, and the initial weights of the additional networks are also randomly initialized. Then, the results show that the standard deviation is small, so the influence of the size of neural networks and initial weights can be excluded.

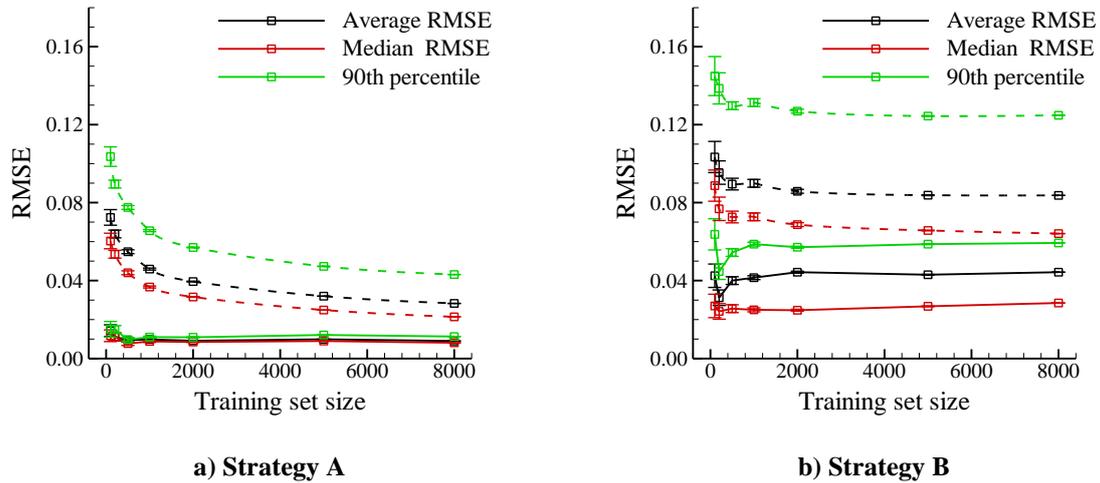

a) Strategy A          b) Strategy B

**Fig. 17 Pressure coefficient distribution prediction errors of different training set (wing section) sizes**
**(Solid: training set; dashed: test set)**

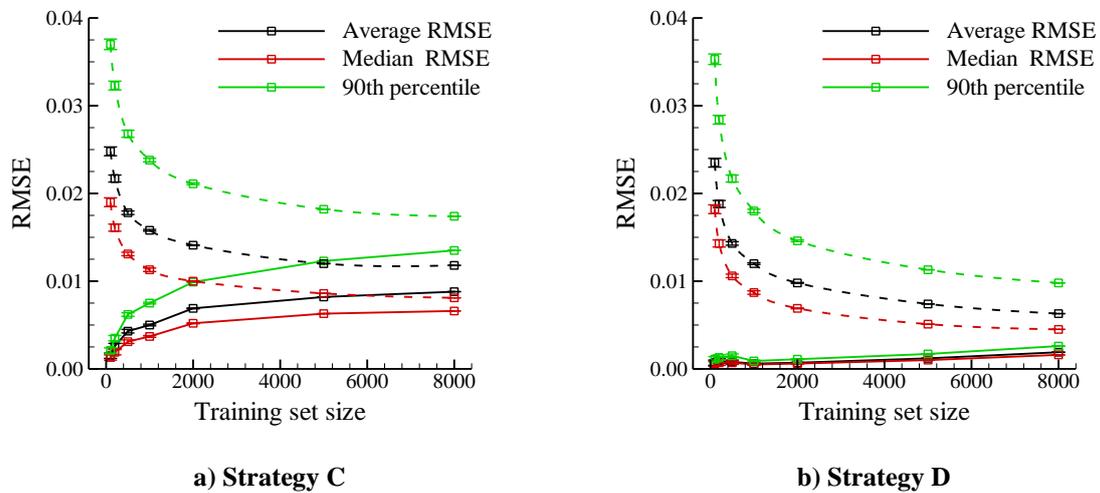

a) Strategy C          b) Strategy D

**Fig. 18 Geometry prediction errors of different training set (wing section) sizes**
**(Solid: training set; dashed: test set)**

Fig. 17 and Fig. 18 show that the four strategies achieve relatively good performance with 2,000 training wing section samples. As introduced in Algorithm A, five sections are extracted from one wing sample, but some of the sections do not meet the constraint listed in Section 2.3. Therefore, 500 wing samples are usually sufficient for the 2D-to-3D transfer learning of the predictive models.

## 4.3 Comparison between the two transfer learning strategies

Fig. 17 and Fig. 18 show that strategies A and D have better performance, i.e., transfer learning achieves better performance when the additional neural network (NN) is appended to the pressure coefficient distribution $\boldsymbol{p}$ than the geometry coefficients $\boldsymbol{g}$.

Fig. 17 a) indicates that transfer learning (strategy A) can achieve an average RMSE of 3.95% for wing section pressure coefficient distribution predictions on the test set. Its 90$^{th}$ percentile of RMSE on the test set is 5.80% (average result of four runs), which is approximately 50% smaller than strategy B. Fig. 19 shows the pressure coefficient distribution examples of the median, 90$^{th}$ percentile and the maximum RMSE on the test set. Since the mean RMSE of the baseline model on the wing sections is 10.30%, as shown in Section 3.2 and Fig. 12, transfer learning can reduce the pressure coefficient distribution prediction error by 61%.

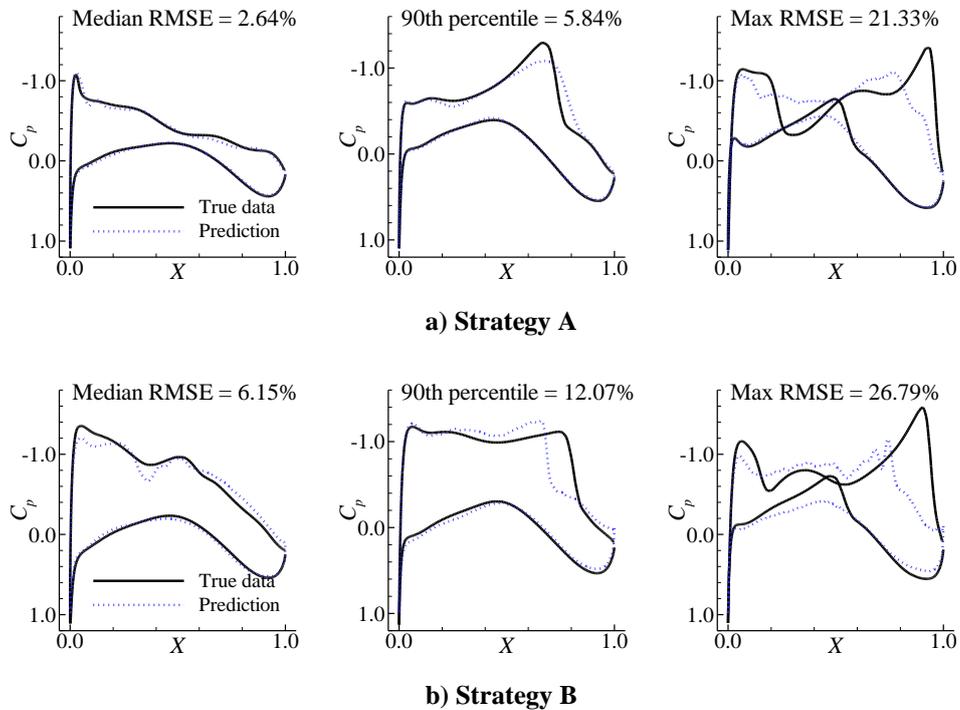

Fig. 19 Examples of pressure coefficient distribution prediction with different strategies

Fig. 18 b) indicates that transfer learning (strategy D) can achieve an average RMSE of 0.98% for wing section geometry predictions on the test set. Its 90$^{th}$ percentile of RMSE on the test set is 1.40% (average result of four runs), which is approximately 30% smaller than strategy C. Fig. 20 shows the geometry prediction examples of the median, 90$^{th}$ percentile and the maximum RMSE on the test set. Since the mean RMSE of the baseline model on the wing sections is 4.94%, as shown in Section 3.2 and Fig. 12, transfer learning can reduce the geometric prediction error by 81%.

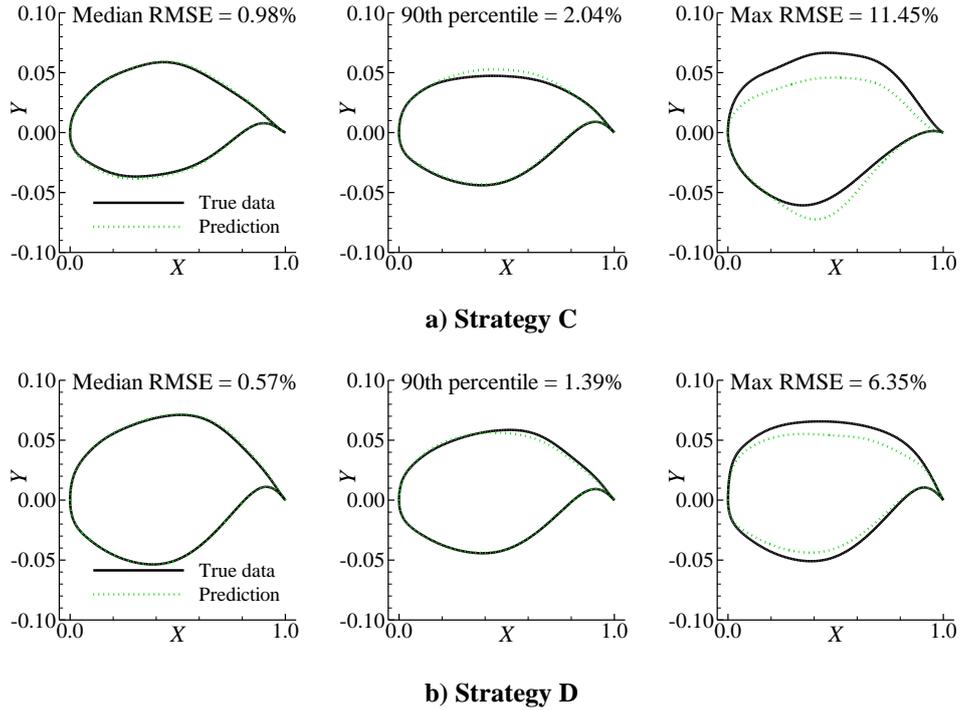

a) Strategy C

b) Strategy D

**Fig. 20 Examples of geometry prediction with different strategies**

The RMSE on the training set slowly increases as the training set size increases. The increment of RMSE in strategies A, B, and D is rather small, except for strategy C. Since the standard deviation of four runs of different neural network depths is small, the increase in RMSE is not caused by underfitting. Therefore, this situation is probably due to the weakness in strategy C and its model architecture. Further study is needed to determine the reason for the poor performance of strategy C, but both strategies A and D can achieve good prediction accuracy with approximately 500 training wing samples.

**4.4 Test of the transferred models on supercritical wings**

In this section, the transferred models of strategies A and D are tested on 3D wings. The two models are denoted by 'Model A' and 'Model D', respectively. As introduced in Section 2.3, there are 500 wing samples that are neither used for the training nor the testing in Section 4.2. As shown in Fig. 6, multiple sections can be used to construct a complete 3D wing. In this paper, 21 sections evenly distributed in the range of 20%-90% semispan are used to construct the wing. The 500 wing samples are further selected to ensure that at least 18 of the 21 sections of the selected wing samples meet the constraints listed in Section 2.3. Then, 200 valid wing samples are selected for the testing of the transferred models.

The RMSE of a wing sample is defined as the average RMSE of its 21 sections. The RMSE is calculated with the transformed pressure coefficient and Y coordinates so that the scale of RMSE remains the same as in Section 4.3. The average RMSE of 200 wings in the pressure coefficient prediction is 3.08%, and the average RMSE of

200 wings in the geometry prediction is 0.68%. Fig. 21 shows some examples of the predicted pressure coefficient contour on different wing planforms (the contour on the upper surface is demonstrated). The prediction value is plotted in the region between the two black lines on the left wing, and the rest of the contour on the aircraft surface is the true pressure coefficient, showing that Model A can predict relatively accurate pressure coefficient contours on various wing planforms and free stream conditions. The spanwise continuity of the prediction is also good.

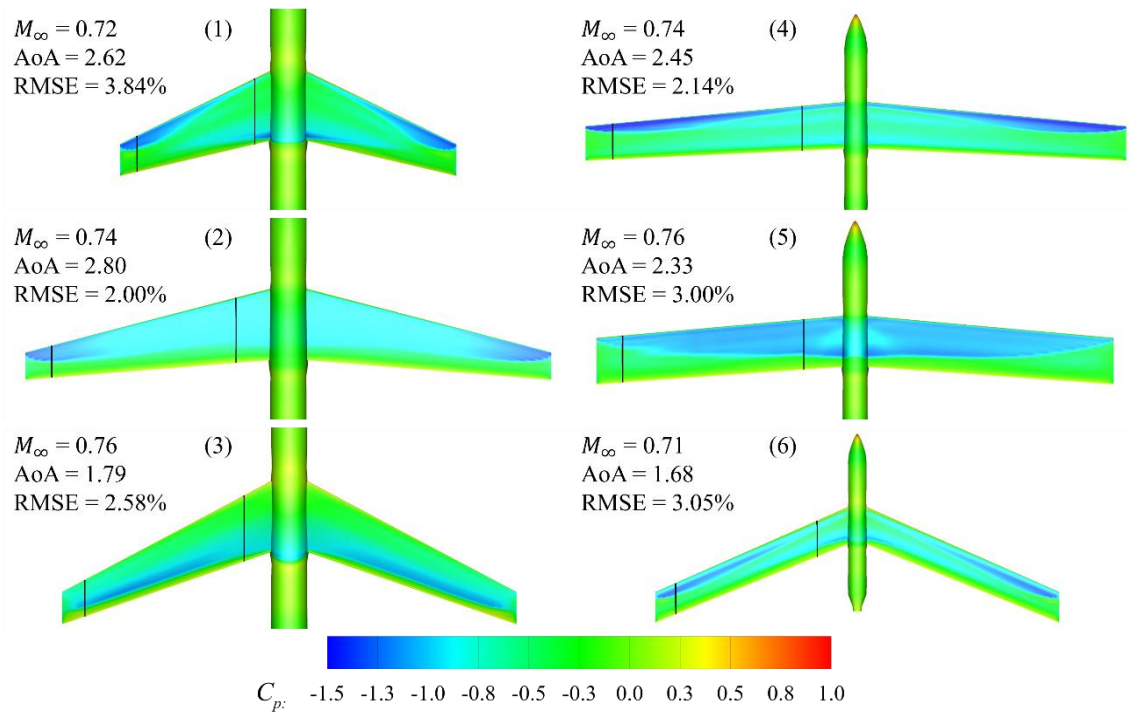

**Fig. 21 Comparison between the true pressure coefficient contour and the predicted contour**
**(Prediction: the contour between the black lines on the left wing; true value: the other contour)**

Fig. 22 shows the prediction error on the six wings in Fig. 21. Fig. 23 shows the pressure coefficient distribution on the half semispan (i.e., 50% $Z_{tip}$) section. The error is the absolute difference between the predicted value and the true data. Fig. 22 indicates that the pressure coefficient contour predicted by Model A sometimes has a large error in the shock wave region, mainly because the gradient is large in these regions, which makes prediction difficult. However, in general, Model A still gives reasonable and relatively accurate predictions.

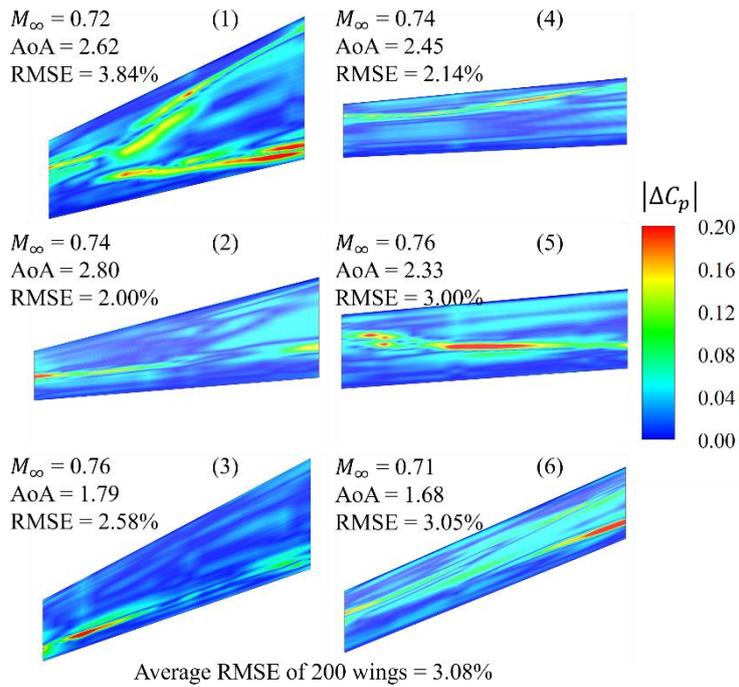

**Fig. 22 Error distribution of the predicted pressure coefficient contour on the upper surface**

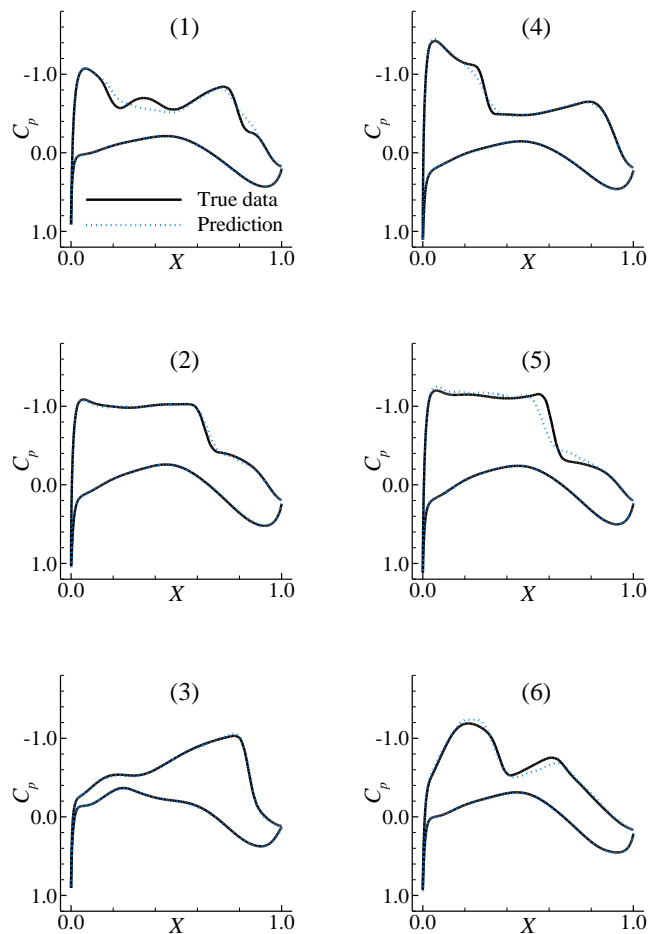

**Fig. 23 Pressure coefficient distribution predictions and true data at the half semispan location**

Fig. 24 shows the prediction error on the six wings in Fig. 21. Fig. 25 shows the wing section geometry at the half semispan (i.e., 50% $Z_{tip}$) location. The error is the absolute difference between the predicted value and the true data. The Y coordinates of the wing surface are smooth; therefore, the prediction error is much smaller for Model D.

In summary, although the transferred model has a relatively large pressure coefficient prediction error in the shock wave regions, the transferred model achieves a good overall prediction accuracy under different wing planforms and free stream conditions. The spanwise continuity and smoothness of the prediction values are good as well.

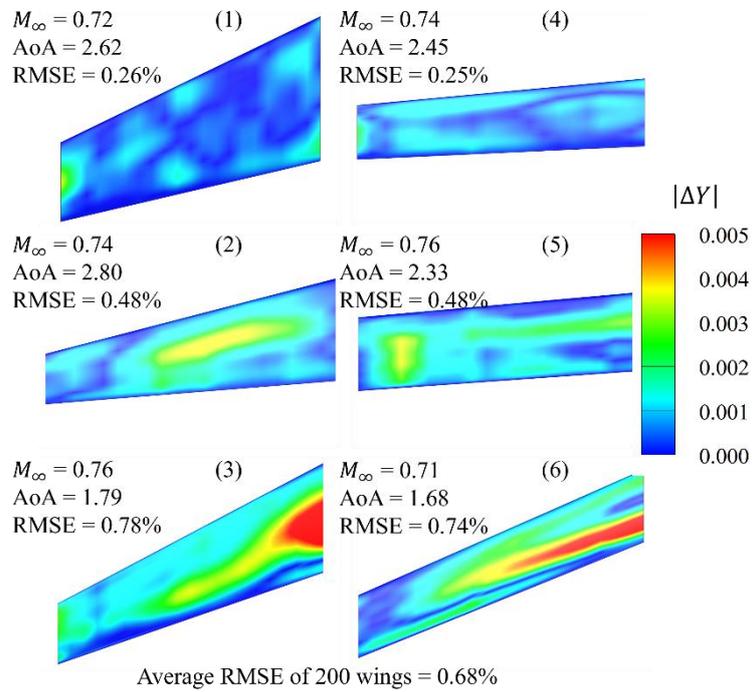

**Fig. 24 Error distribution of the predicted Y coordinates of the upper surface**

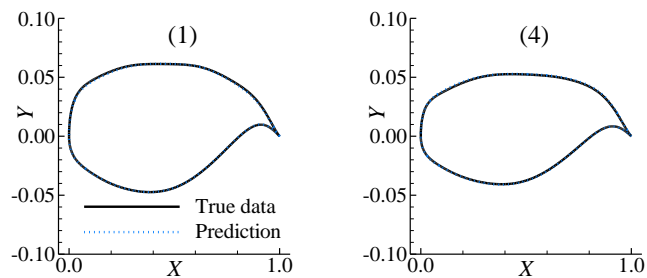

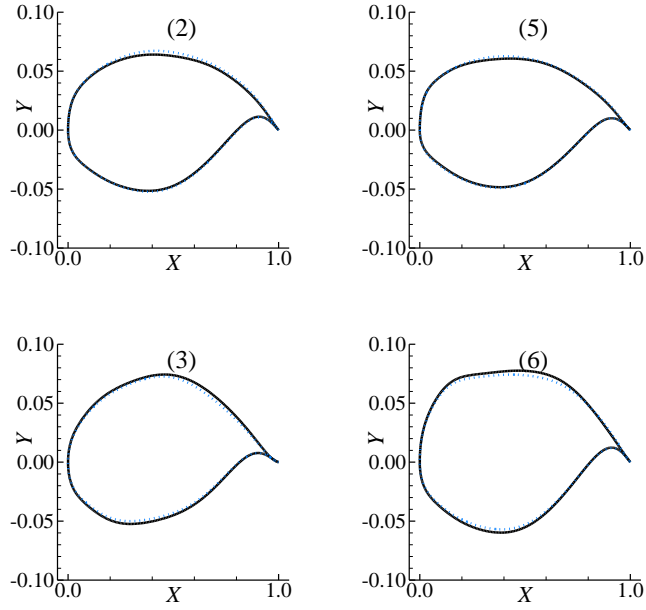

**Fig. 25 Geometry predictions and true data at the half semispan location**

## V. Conclusions

Transfer learning is a popular approach to reuse a pretrained model on a different but related problem, which often greatly reduces the need for data. This paper proposes to use transfer learning to reuse machine learning models trained on two-dimensional supercritical airfoils so that these baseline models can be used on finite-span swept wings. Two models for the forward problem and inverse problem in fluid mechanics are trained as the baseline models for transfer learning. The forward problem model predicts the pressure coefficient distribution based on the geometry, and the inverse problem model predicts the geometry based on the pressure coefficient distribution.

Wing samples with different planform parameters and free stream conditions are randomly sampled. Multiple wing sections are randomly extracted from the wings; then, the geometry and pressure coefficient distribution are transformed to the equivalent two-dimensional airfoils by the simple swept theory for training. Two transfer learning strategies are applied to two baseline models. The difference between the two strategies and different training set sizes is studied. The following conclusions are reached.

(1) The simple swept theory is embedded in the parameter-based transfer learning process to describe the general influence of the swept angle. The parameters of the baseline models are fixed so that the knowledge contained in the baseline models can be transferred. Then, an additional neural network (NN) is appended to the baseline models and trained to capture the remaining 3D effects caused by the finite span, spanwise geometry change (e.g., tapper ratio, thickness change), and wing-body interference.

(2) Transfer learning can reduce the prediction error by 61% compared to the baseline model that predicts the pressure coefficient distribution of wing sections and can also reduce the prediction error by 81% compared to the baseline model that predicts the geometry of wing sections.

(3) The transferred Model A achieves an average RMSE of 3.08% on the prediction of 200 wing pressure coefficient contours. The transferred Model D achieves an average RMSE of 0.68% on the prediction of 200 wing geometries. Although Model A has a relatively large prediction error in the shock wave regions, the overall prediction accuracy is acceptable for most wing planforms and free stream conditions.

(4) The results indicate that it is better to append the additional neural network (NN) to the pressure coefficient distribution $p$ than the geometry coefficients $g$ in parameter-based transfer learning. In addition, 500 wing samples are sufficient for the 2D-to-3D transfer learning of the baseline models.

In summary, transfer learning is a promising approach to save computational costs on building models for three-dimensional wings. However, more complex wing geometries and aircraft configurations await further study.

## Acknowledgments

This work was supported by the National Natural Science Foundation of China under Grant Nos. 92052203, 91852108 and 11872230.This work was supported by the National Natural Science Foundation of China under Grant Nos. 92052203, 91852108 and 11872230.